\documentclass[preprint,showpacs,preprintnumbers,amsmath,amssymb,natbib]{revtex4}

\usepackage{graphicx}
\usepackage{epsfig}		
\usepackage{dcolumn}
\usepackage{bm}
\def\0{\mbox{\tiny $0$}}
\def\1{\mbox{\tiny $1$}}
\def\2{\mbox{\tiny $2$}}
\def\3{\mbox{\tiny $3$}}
\def\4{\mbox{\tiny $4$}}
\def\5{\mbox{\tiny $5$}}
\def\6{\mbox{\tiny $6$}}
\def\7{\mbox{\tiny $7$}}
\def\8{\mbox{\tiny $8$}}
\def\9{\mbox{\tiny $9$}}

\def\f14{\mbox{\tiny $\frac{1}{4}$}}

\DeclareMathOperator{\sech}{sech}

\begin{document}

\title{Charge varying sine-Gordon deformed defects}

\author{A. E. Bernardini}
\email{alexeb@ufscar.br}
\author{M. Chinaglia}
\email{chinaglia.mariana@gmail.com}
\affiliation{Departamento de F\'{\i}sica, Universidade Federal de S\~ao Carlos, PO Box 676, 13565-905, S\~ao Carlos, SP, Brasil}
\author{Rold\~ao da Rocha}
\email{roldao.rocha@ufabc.edu.br}
\affiliation{Centro de Matem\'atica, Computa\c c\~ao e Cogni\c c\~ao,
Universidade Federal do ABC 09210-170, Santo Andr\'e, SP, Brazil}
\date{\today}

\begin{abstract}
Sine-Gordon deformed defects that exhibit unusual phenomenological features on the topological charge are investigated.
The possibility of a smooth and continuous transition between topological (non null charge) and non-topological (null charge) scenarios of deformed defects supported by sine-Gordon structures is evinced by the analytical calculation of topological charges and localized energy distributions.
By describing cyclic deformation chains, we show that a triggering sine-Gordon model simultaneously supports kink and lump-like defects, whose topological mass values are closed by trigonometric or hyperbolic successive deformations.
In spite of preserving analytical closure relations constraining the topological masses of $3$-and $4$-cyclically deformed defects, the deformation chains produce kinks and lumps which exhibit nonmonotonic behavior and extra inflection points.
The outcome of our analysis suggests that cyclic deformations create novel scenarios of physical and mathematical applicability of defect structures supported by the sine-Gordon theory.
\end{abstract}

\pacs{05.45.Yv, 03.65.Vf, 11.27.+d}
\keywords{sine Gordon - deformed defects - lumps - kinks}
\date{\today}
\maketitle

\section{Introduction}

Among nonlinear systems, the sine-Gordon (SG) model \cite{BookB,BookB2,SG01,SG02,SG03,SG04,SG05} supports a wide range of reshaped and deformed topological models \cite{BookA,BookA22,0001,0002,0003,0004,000422,BSG} that exhibit an ample scenario of topological (kink-like) and non-topological (lump-like) analytical solutions, with theoretical and phenomenological applications in several fields \cite{BookA,BookA22,Books,BooksAAA,BooksBBB}.
From the point of view of its inherent symmetry properties, the SG model solutions exhibit Lorentz invariance and finite topological charges represented by kink, anti-kink and breather solutions \cite{BookB,BookB2}.
In addition, it has been pointed out that the quantum integrability described by the Yang-Baxter equation for the SG scenario leads to the quantum algebra $su_q(2)$ \cite{Faddeev,Sklyanin}, and in string scenarios the SG model is supposed to be related to the classical string on some particular manifolds as well \cite{String,Mikhailov,String2}.

Notwithstanding the ferment in its subjacent mathematical field, the SG theory is investigated in many contexts of applied physics \cite{Books, BooksAAA, BooksBBB,App01,App01AAA,App01BBB,App01CCC,App02,App02AAA,App02BBB,App03}.
SG models with variable physical mass appear, for instance, in a description of media inhomogeneities of Josephson junctions \cite{App01,App01AAA,Josephson01,Goldobin,Susanto} as well as in the study of molecular \cite{Books, BooksAAA, BooksBBB} and DNA-promoter dynamics \cite{DNA,DNA2}.
Likewise, reshaped lump-like models have been considered in investigating the behavior of bright solitons in optical fibers \cite{0017A,0018A}, by the way, a distinguishable alternative for obtaining ultra-high-speed packet-switched optical networks.

Given that obtaining (non)topological defect structure solutions circumstantially requires novel and creative mathematical devices that support analytical protocols, we investigate a set of deformation procedures \cite{0004,000422,Bas01,Bas04,Bas02} for models with well-defined topological masses and charges derived from cyclically deformed defects \cite{AlexRoldao}, which are constrained by closure relation involving their topological masses.
Deformed defects are generated and regenerated from a primitive defect of the SG theory for a $1+1$ dimension scalar field, $\chi$.

As well known, for SG kink-like structures, the topological defect corresponds to an interface between two regions of a scalar field potential with different minima.
Described in terms of a spatial coordinate, $s$, it is identified by a topological charge given by
\begin{equation}
Q=\chi(s\rightarrow + \infty)-\chi(s\rightarrow -\infty),
\end{equation}
(as it currently defined in elementary textbooks, for instance, through Eq.~(2.58) on pag. 33 from Ref.~\cite{BookA,BookA22}, and through Eq.~(3.6) on pag. 198 from Ref.~\cite{Col85})

These defects are therefore defined by localized energy densities supported by solutions of a nonlinear partial differential equation, which tend to be stabilized by such a conserved charge which indeed is supported by an underlying field theory \cite{BookB}.

Systematic modifications of the SG model due to cyclic deformation chains comprise novel analytically solvable SG-like scenarios for which a topological mass (corresponding to a finite energy profile) and the topological charge can be analytically obtained.
Such a deformation framework also supports the generation of lump-like solutions, namely the non-topological ones, for which the topological charge vanishes, i. e. $Q \sim 0$.
The main feature of the deformed SG models here investigated is the phenomenological parametrization of a smooth and continuous transition between topological ($Q \neq 0$) and non-topological ($Q = 0$) scenarios.

Although our work has been concerned with the non constant profile of the topological charge, $Q_{(n)}$, namely a charge varying behavior, which is driven by a phenomenological parameter, $n$, the topological masses \cite{BookB,BookB2} are also obtained for novel cyclic scenarios of deformed models supported by a SG generating model.
As a matter of convention \cite{AlexRoldao}, the nomenclature of topological mass hereon is extended to encompass any prescription of finite total energy obtained through the integration of a field dependent localized energy density, $\rho(\phi)$ \cite{BookB} as
\begin{equation}
M = 
\left|\int^{+\infty}_{-\infty}d s \,\rho(\phi(s))\right|,
\end{equation}
even for those of non-topological lump-like structures.

The outline of this manuscript is as follows.
In section II we report about the framework of $N\hspace{-.1 cm}+\hspace{-.1 cm}2$-cyclic deformation procedures \cite{AlexRoldao} and their eventual extensions to SG models.
The explicit analytical solutions for deformed SG models built from $3$ and $4$-cyclic deformation chains are presented in sections III and IV, respectively, where we find that the deformed solutions seem to support some exotic two-kink solutions and kink and lump-like structure phenomenologically connected by a free parameter, $n$.
Some peculiarities on the results from sections III and IV are also quantified in terms of the behavior of the variable topological masses and charges which are analytically obtained.
Finally, our conclusions are drawn in section V.

\section{Cyclic deformation}

Departing from a triggering primitive defect, $\chi$, which might be identified by the SG kink solution, one is able to construct an $N\hspace{-.1 cm}+\hspace{-.1 cm}2$-cyclic deformation chain through the application of a chain rule constrained by hyperbolic and trigonometric fundamental relations \cite{AlexRoldao}.

The generalized $\lambda$-deformation operation, with $\lambda =0,\, 1,\,2,\, \ldots,\,N$  is effectuated by the $\lambda$-derivative given by
\begin{equation}
g^{[\lambda]}(\phi^{[\lambda]}) = \frac{d\phi^{[\lambda]}}{d\phi^{[\lambda-1]}},
\end{equation}
where $\phi^{[\lambda]}$ are real scalar fields corresponding to defect structures built from an $N\hspace{-.1 cm}+\hspace{-.1 cm}2$-cyclic deformation chain triggered by $\chi\equiv\chi(s)\sim \phi^{[-1]}$, such that one effectively has $\phi^{[\lambda]} \equiv \phi^{[\lambda]}(\chi)$.

The hyperbolic deformation chain \cite{AlexRoldao} is given by
\begin{eqnarray}
\phi^{[0]}_{\chi} &=& \tanh{(\chi)}, \nonumber\\
\phi^{[1]}_{\chi} &=& \tanh{(\chi)}\sech{(\chi)}, \nonumber\\
\phi^{[2]}_{\chi} &=& \tanh{(\chi)}\sech{(\chi)}^{2}, \nonumber\\
&\vdots& \nonumber\\
\phi^{[N-1]}_{\chi} &=& \tanh{(\chi)}\sech{(\chi)}^{N-1}, \nonumber\\
\phi^{[N]}_{\chi} &=& \sech{(\chi)}^{N},
\label{AA}
\end{eqnarray}
where the subindex denotes the corresponding derivative, which upon straightforward integrations leads to
\begin{eqnarray}
\phi^{[0]}(\chi) &=& \ln{[\cosh{(\chi)}]}, \nonumber\\
\phi^{[1]}(\chi) &=& - \sech{(\chi)}, \nonumber\\
\phi^{[2]}(\chi) &=& - \frac{1}{2} \sech{(\chi)}^{2}, \nonumber\\
&\vdots& \nonumber\\
\phi^{[N-1]}(\chi) &=& - \frac{1}{N-1}\sech{(\chi)}^{N-1}, \nonumber\\
\phi^{[N]}(\chi) &=&  \sinh{(\chi)}\, _2F_1\left[\frac{1}{2},\, \frac{1+N}{2},\, \frac{3}{2},\, -\sinh{(\chi)}^2\right], \nonumber\\
\label{BB}
\end{eqnarray}
where $_2F_1$ is the Gauss' hypergeometric function and the arbitrary integration constants have been suppressed, since they are not effective in this framework.

From Eqs.~(\ref{AA}-\ref{BB}) one identifies
\begin{eqnarray}
g^{[\lambda]}(\phi^{[\lambda]}) = \sech{(\chi)} \equiv \exp{[-\phi^{[0]}]},&~~& \lambda = 1,\,2,\, \ldots,\,N-1,\\
g^{[N]}(\phi^{[N]}) = 1/\sinh{(\chi)} ,&~~&
\end{eqnarray}
and
\begin{equation}
\prod_{\lambda=1}^{N-1}{g^{[\lambda]}(\phi^{[\lambda]})} = \frac{d\phi^{[N-1]}}{d\phi^{[0]}} = \sech{(\chi)}^{N-1} \equiv \exp{[- (N-1) \phi^{[0]}]},
\label{DD}
\end{equation}
from which the complete expression for the chain rule of the $N\hspace{-.1 cm}+\hspace{-.1 cm}2$-cyclic deformation can be written as
\begin{equation}
\frac{d\phi^{[0]}}{d\chi}\,\prod_{\lambda=1}^{N}{g^{[\lambda]}(\phi^{[\lambda]})}\,\frac{d\chi}{d\phi^{[N]}} = 1,
\label{EE}
\end{equation}
from which $N\hspace{-.1 cm}+\hspace{-.1 cm}2$ deformation functions closing the cycles can be identified.

By assuming a parametrization in terms of generalized BPS functions \cite{BPS,BPS2}, one has
\begin{eqnarray}
y^{[\lambda]}_{\phi^{[\lambda]}} &=& \frac{d \phi^{[\lambda]}}{ds} = y^{[\lambda-1]}_{\phi^{[\lambda-1]}}\, g^{[\lambda]}(\phi^{[\lambda]}) =
y^{[\lambda-r]}_{\phi^{\lambda-r]}}\, g^{[\lambda-r+1]}(\phi^{[\lambda-r+1]})\nonumber\\
&=& w_{\chi} \phi^{[\lambda]}_{\chi},
\label{nove}
\end{eqnarray}
with $w_{\chi} = d\chi/ds$.
In this case, the constraint equation is written as
\begin{eqnarray}
\sum_{\lambda=0}^{N-1}{(y^{[\lambda]}_{\phi^{[\lambda]}})^2}
&=& (y^{[0]}_{\phi^{[0]}})^2 \sum_{\lambda=0}^{N-1}{\left(\frac{d\phi^{[\lambda]}}{d\phi^{[0]}}\right)^2} \nonumber\\
&=&  w_{\chi}^2 \sum_{\lambda=0}^{N-1}{(\phi^{[\lambda]}_{\chi})^2}
= w_{\chi}^2 \tanh{(\chi)}^2\,\sum_{\lambda=0}^{N-1}{(\sech{(\chi)})^{2\lambda}}\nonumber\\
&=&  w_{\chi}^2 \tanh{(\chi)}^2\,\frac{1 - \sech{(\chi)}^{2N}}{1-\sech{(\chi)}^2} = w_{\chi}^2 \left( 1 - \sech{(\chi)}^{2N}\right)\nonumber\\
&=& w_{\chi}^2 \left[ 1 - (\phi^{[N]}_{\chi})^2 \right] = w_{\chi}^2 - (y^{[N]}_{\phi^{[N]}})^2,
\label{BBBB}
\end{eqnarray}
which leads to
\begin{eqnarray}
\sum_{\lambda=0}^{N}{(y^{[\lambda]}_{\phi^{[\lambda]}})^2} &=& w_{\chi}^2.
\label{cd1}
\end{eqnarray}

Therefore, the preliminary relevant result reported on \cite{AlexRoldao} is that the topological masses defined through \cite{BookB,BookB2,Col85}
\begin{equation}
M^{\phi^{[\lambda]}} = 
\left|\int^{+\infty}_{-\infty}{d s \,(y^{[\lambda]}_{\phi^{[\lambda]}})^2}\right| 
\end{equation}
and
\begin{equation}
M^{\chi} = \left|\int^{+\infty}_{-\infty}{ds \, (w_{\chi})^2}\right| 
\end{equation}
follow the constraint given by
\begin{eqnarray}
\sum_{\lambda=0}^{N}{(M^{\phi^{[\lambda]}})} &=& M^{\chi},
\label{cd2}
\end{eqnarray}
where the corresponding localized energy densities have been identified with
\begin{equation}
\rho(\phi) = y_{\phi}^2 \qquad \mbox{and} 
\qquad 
\rho(\chi) = w_{\chi}^2
\end{equation}
as to identify $M$ with finite energy configurations of localized structures (in this case, of sine-Gordon models).

Following the above framework, also a trigonometric deformation chain \cite{AlexRoldao} can be obtained by concomitantly changing $\tanh{(\chi)}\mapsto-\sin{(\chi)}$ and
$\sech{(\chi)}\mapsto\cos{(\chi)}$ into Eqs.~(\ref{AA}-\ref{BB}), such that
\begin{eqnarray}
\phi^{[0]}(\chi) &=& \cos{(\chi)}, \nonumber\\
\phi^{[1]}(\chi) &=& \frac{1}{2}\cos{(\chi)}^{2}, \nonumber\\
\phi^{[2]}(\chi) &=& \frac{1}{3}\cos{(\chi)}^{3}, \nonumber\\
&\vdots& \nonumber\\
\phi^{[N-1]}(\chi) &=& \frac{1}{N}\cos{(\chi)}^{N}, \nonumber\\
\phi^{[N]}(\chi) &=&  -\frac{\cos{(\chi)^{N+1}}}{N+1}\, _2F_1\left[\frac{1+N}{2},\, \frac{1}{2},\, \frac{3 + N}{2},\, \cos{(\chi)}^2\right].
\label{BB2}
\end{eqnarray}
It results into a set of constraints analogous to those described by Eqs.~(\ref{cd1}-\ref{cd2}).

Since $N\hspace{-.1 cm}+\hspace{-.1 cm}2$-cyclic deformation with the above-obtained topological mass constraints can be systematically constructed in terms of bijective trigonometric and hyperbolic functions, one can describe all deformed SG models by assuming that $\chi(s)$ is supported by
a SG theory whose potential, $T(\chi)$, is given by
\begin{equation}
\label{potphi4}
T(\chi)=1-\cos(\chi) = 2\sin\left(\frac{\chi}{2}\right)^{2} = \frac{1}{2} w_{\chi}^2,
\end{equation}
as to have
$$w_{\chi} = 2\sin\left(\frac{\chi}{2}\right),$$
as an input into Eq.~(\ref{nove}),
where it is assumed that all the dimensional constants have been absorbed by the scalar field, $\chi$, which is in a dimensionless representation \cite{BookB,BookB2}.

Solving the equation of motion in an also dimensionless representation of the dynamical quantities \cite{BookB}, $t$ and $s$,
\begin{equation}
\frac{\partial^2{\chi}}{\partial t^2}-\frac{\partial^2{\chi}}{\partial s^2}+\sin(\chi) = 0,
\label{eqsine}
\end{equation}
one obtains the topological static solution (see, for instance, Eqs.~(2.63-2.68) on pages 35-36 from Ref.~\cite{BookB} and the {\em Gudermanian function} definition from Section 4.23 of Ref.~(\cite{Olv})),
\begin{equation}
\chi(s)=\pm 4 \arctan{\left(e^{s}\right)}-\pi,
\label{solsine}
\end{equation}
where the relative constant/coefficients are introduced in order to keep the choice for (anti)symmetric representations of the defects.

Finally,by substituting the above primitive solution as well as the derivatives related to Eq.~(\ref{BB2}) into Eq.~(\ref{nove}) one obtains
\begin{eqnarray}
y^{[N-1]}_{\phi^{[\lambda]}} &=& w_{\chi} \phi^{[N-1]}_{\chi} = - 2 \sin\left(\frac{\chi}{2}\right)\sin(\chi) \,\cos(\chi) ^{(N-1)}
\end{eqnarray}
from which one is constraint to take advantage of the cyclic chain to obtain the analytical form of the defects.
Although one could obtain an enormous set of cyclically deformed SG defect structures,
our following results are constrained by those which guarantee the existence of simplified closure relations involving the topological masses and (semi)analytical expressions for the topological charges of the deformed defects.
In particular, we shall consider only $3$- and $4$-cyclic chain models triggered by the kink-like solution, $\chi(s)$.

\section{$3$-cyclic SG models}

In a $3$-cyclic chain, the scalar field, $\chi$, henceforth will designate the primitive kink defect so that one has
\begin{eqnarray}
w_{\chi}    &=& \frac{d \chi}{d s} = y_{\phi} \chi_{\phi} = z_{\psi} \chi_{\psi},
\label{topo023}
\end{eqnarray}
with
\begin{eqnarray}
y_{\phi}    &=& w_{\chi} \phi_{\chi}, \nonumber\\
z_{\psi}    &=& w_{\chi} \psi_{\chi},
\label{topo023B}
\end{eqnarray}
and the cyclically derived BPS potentials, which belong to the $3$-cyclic deformation chain, are then given by
\begin{equation}
T(\chi) = \frac{1}{2} w_{\chi}^2 ~\rightleftarrows ~
V(\phi) = \frac{1}{2} y_{\phi}^2 ~\rightleftarrows ~
W(\psi) = \frac{1}{2} z_{\psi}^2 ~\rightleftarrows ~
T(\chi) = \frac{1}{2} w_{\chi}^2,
\label{topo023C}
\end{equation}
with three deformation functions constrained by the chain rule,
\begin{equation}
\psi_{\phi} \phi_{\chi} \chi_{\psi}= 1,
\label{topo033}
\end{equation}
where the SG solution (\ref{solsine}) leads to
\begin{equation}
\omega_{\chi}=\frac{d\chi}{ds}=\frac{4e^{s}}{1+e^{2s}}.
\end{equation}

\subsection{Hyperbolic deformation}

Let us consider the set of auxiliary derivatives described by
\begin{eqnarray}
\label{funcdeformhypsine}
\phi^{(n)}_{\chi} =  \sech[(\chi-(n-1)\pi)],\nonumber\\
\psi^{(n)}_{\chi} = -\tanh[(\chi-(n-1)\pi)],
\label{hyp03}
\end{eqnarray}
which upon straightforward integrations lead to
\begin{eqnarray}
\label{defdeformadoshyp}
\phi(\chi,n) = 2 \arctan[\tanh[(\chi - (n - 1) \pi)/2]],\nonumber\\
\psi(\chi,n) = -\ln[\cosh[ \chi - (n - 1) \pi]/\cosh(\pi)],
\label{hyp3}
\end{eqnarray}
with constants and overall signs chosen to fit ordinary values for the asymptotic limits.
After simple mathematical manipulations involving the hyperbolic fundamental relation $\tanh{(n\,\chi)}^2 + \sech{(n\,\chi)}^2 = 1$ and by 
using Eq.~(\ref{topo023B}), one straightforwardly  identifies the following closure relation:
\begin{eqnarray}
w_{\chi}^{2}
&=& w_{\chi}^{2}\left[\tanh{(n\,\chi)}^2 + \sech{(n\,\chi)}^2\right]\nonumber\\
&=& w_{\chi}^{2}\left[\psi^{(n)\,2}_{\chi} + \phi^{(n) \, 2}_{\chi}\right]\nonumber\\
&=& z_{\psi}^{2} + y_{\phi}^{2}\,.
\label{hyp013}
\end{eqnarray}
It constrains the localized energy distributions for $3$-cyclic deformed defect structures.
Eq.~(\ref{hyp013}) corresponds to Eq.~(\ref{cd1}) by setting $N = 1$.

The corresponding BPS deformed functions are thus obtained as
\begin{eqnarray}
y_{\phi}=\omega_{\chi} \sech[\chi-(n-1) \pi],\nonumber\\
z_{\psi}=-\omega_{\chi}\tanh[\chi-(n-1) \pi],
\label{hyp003}
\end{eqnarray}
with the respective dependencies on $\phi^{(n)}(s)$ and $\psi^{(n)}(s)$ obtained by substituting $\chi = \chi(s)$ from Eq.~(\ref{solsine}) into Eq.~(\ref{hyp3}).

\begin{figure}[h!]
\begin{center}
\hspace{-1 cm}
\includegraphics[scale=0.5]{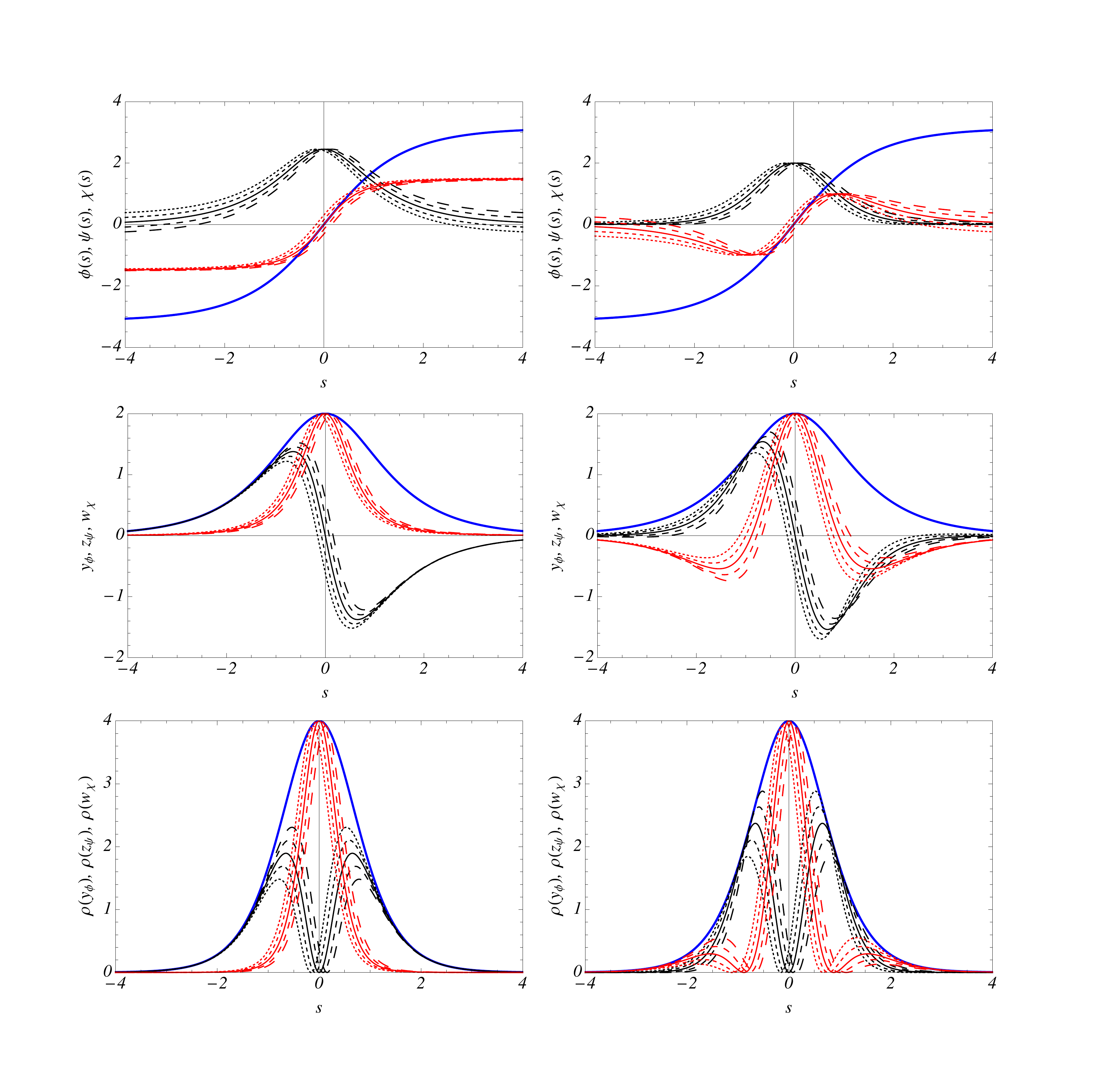}
\end{center}
\caption{(Color online) The $3$-cyclically deformed defects systematically obtained from hyperbolic (first column) and trigonometric (second column) deformation functions.
Results are for the primitive SG kink solution, $\chi(s)$ (thick blue line), and for deformed defects, $\phi^{(n)}(s)$ (red lines) and $\psi^{(n)}(s)$ (black lines).
The hyperbolic(trigonometric) deformation chain leads to kink(lump)-like defects.
On the first row one has the defect profile, on the second row one has the BPS function and on the last row one has the energy density profile.
The values of the free parameter $n$ are picked-up from the interval between $0.9$ (dotted lines) and $1.1$ (long dashed lines) with steps equal to $0.05$ for both hyperbolic and trigonometric chains.
Solid lines correspond to $n = 1$.}
\label{Fig03A}
\end{figure}

\begin{figure}[h!]
\hspace{-1 cm}
\includegraphics[scale=0.5]{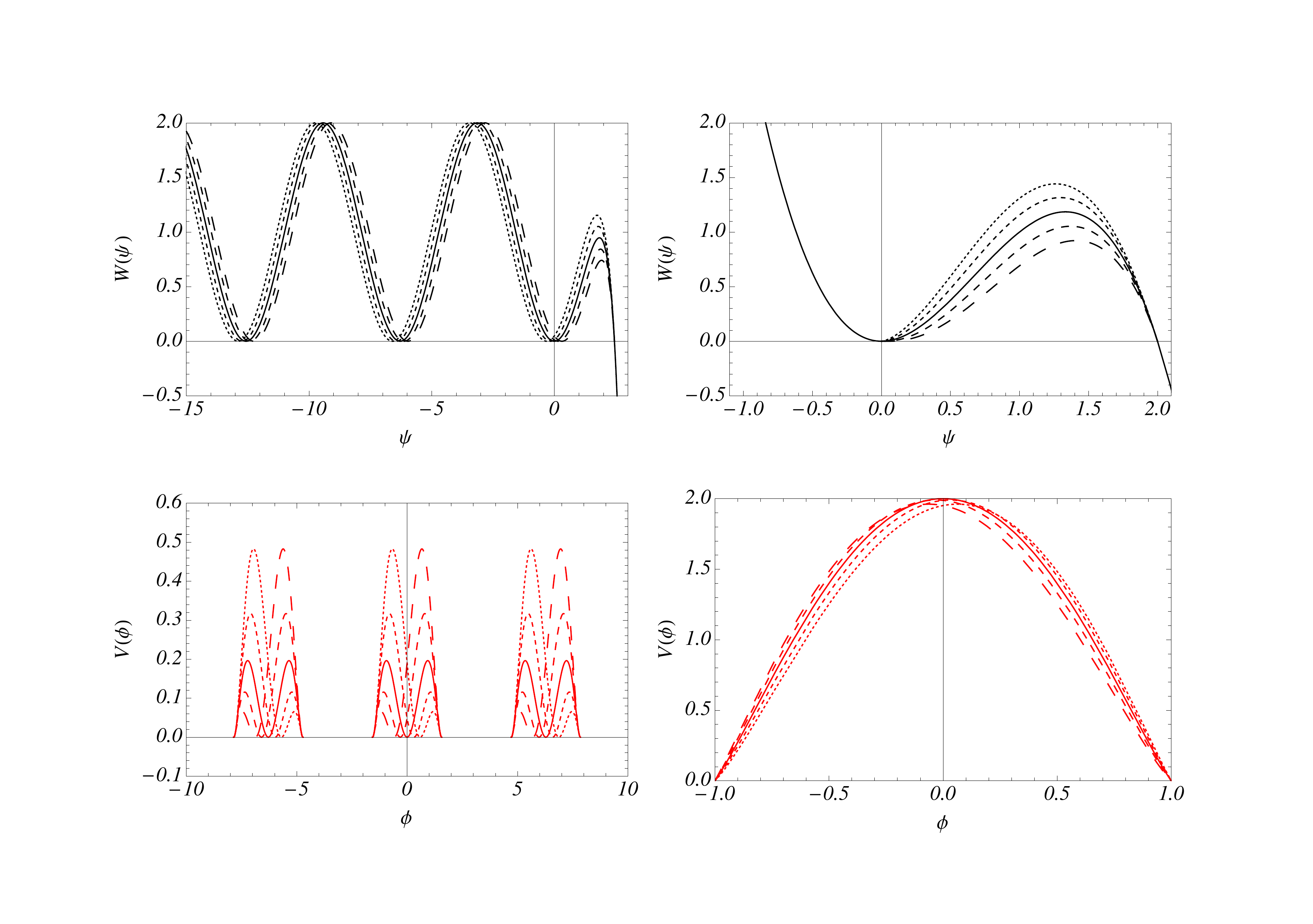}
\caption{(Color online) Deformed potentials, $V(\phi)$ e $W(\psi)$, obtained from the SG model, for hyperbolic (first column) and trigonometric (second column) $3$-cyclic deformation chains.
We have selected the same values of $n$ and we have set the {\em color scheme} correspondence  from Fig.~\ref{Fig03A}.}
\label{Fig03B}
\end{figure}

The first column of Fig.~\ref{Fig03A} shows the analytical dependence on $s$ for the $3$-cyclically deformed defects obtained through the hyperbolic deformation functions from Eq.~(\ref{hyp3}).
The plots show the results for the primitive defects, $\chi(s)$, $w_{\chi}(s) = d\chi/ds$, and $\rho(\chi(s))$; and for the corresponding deformed defects, $\phi^{(n)}(s)$, $y_{\phi}(s) = d\phi^{(n)}/ds$, and $\rho(\phi^{(n)}(s))$; and $\psi^{(n)}(s)$, $z_{\psi}(s) = d\psi^{(n)}/ds$, and $\rho(\psi^{(n)}(s))$. We have set $n = 1 - 0.05k$ with $k\in[-2,2]$, in order to depict the analytical dependence on the free parameter $n$.
The corresponding BPS potentials, $W(\psi)$ and $V(\phi)$, for the same set of $n$ values, are depicted from the first column of Fig.~\ref{Fig03B}.

For the primitive defect engendered by the dimensionless SG theory, i. e. Eq.~(\ref{solsine}), one has both kink- and lump-like defects for given $\psi^{(n)}$.
Such a degenerated behavior is due to some $W(\psi)$ potential features.

For $n=1$ one identifies the unique $\psi$ lump that corresponds to the right part of the $W(\psi)$ curve where the potential shows one minimum and one boundary point.
For all the other values of $n$ the reference frame is moved to the left on the plot (c. f. Fig.~\ref{Fig03B}) so that one shall have a potential that has two minima.
This is the reason for which one has kink-like defects for all other values of $n$. One also notices that the deformed defects bring up one more different feature: the {\em pseudo}lump solutions given by $\psi^{(n)}$ (black lines in Fig.~\ref{Fig03A}) have a nonmonotonic behavior like a lump, but it differs from that because $$\lim_{s\rightarrow- \infty} \psi^{(n)}(s) \neq \lim_{s \rightarrow + \infty} \psi^{(n)}(s),$$ which constrains one to identify $\psi^{(n)}$ as kink solutions.

Analyzing the potential $W(\psi)$, one sees that besides generating an infinite set of critical points for
\begin{equation}
s=0-k\pi,
\end{equation}
with $k=1,\,2,\,3,\,\ldots$, it also generates a boundary point located at $s = -\ln[\sech(\pi)]$.
This point is concerned with the lump obtained for $n=1$.
Therefore one finds a null value for the topological charge for the defect $\psi^{(1)}$ .
The topological charges for $n \neq 1$ are given by:
\begin{equation}
Q^{\psi}_{(n)}=\left\vert\ln\left[\frac{\sech[(n-2)\pi]}{\sech(n\pi)}\right]\right\vert.
\end{equation}

\begin{figure}[h!]
\centering
\includegraphics[scale=0.6]{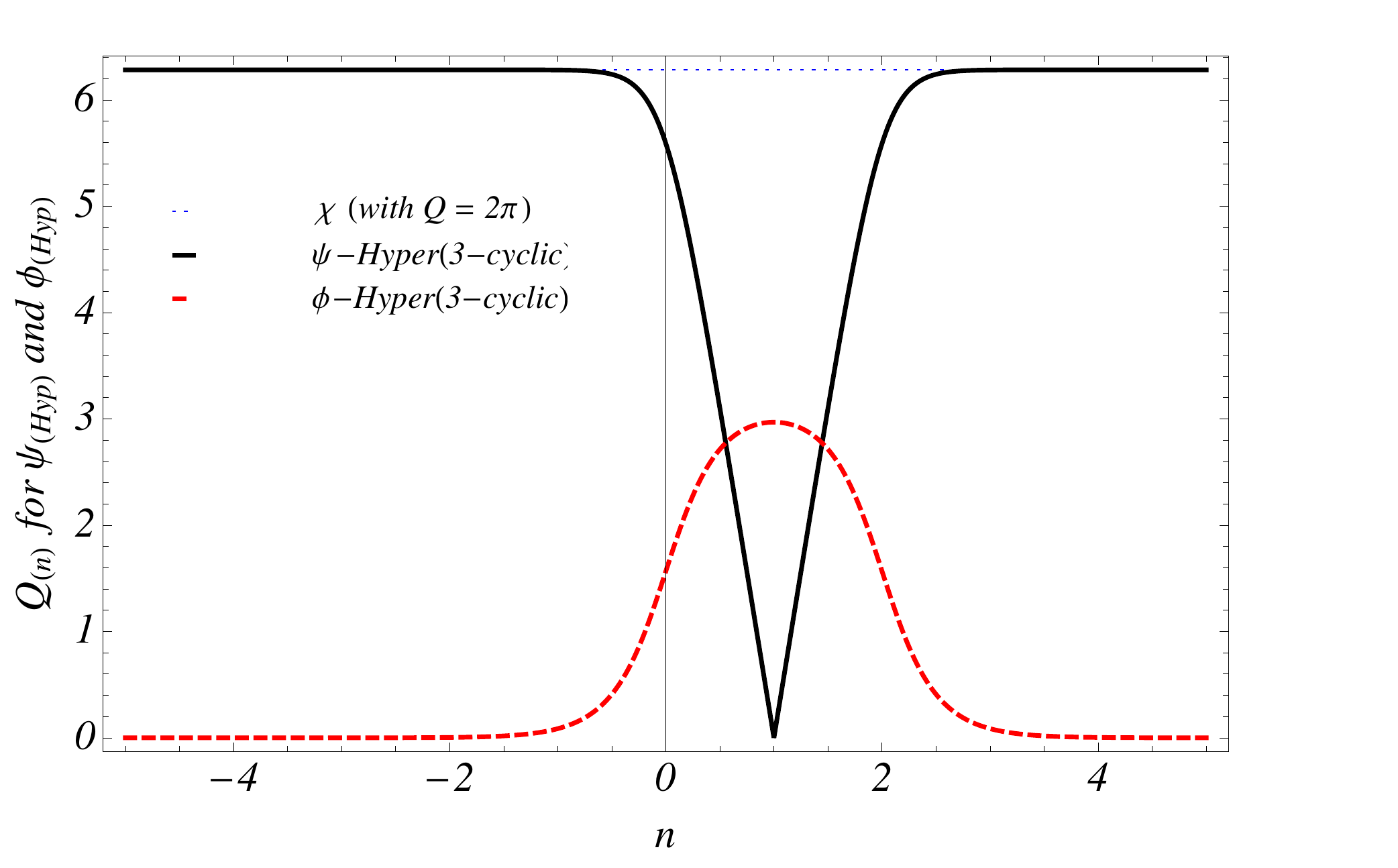}
\caption{(Color online) Topological charges, $Q_{(n)}$, for kink solutions as function of the parameter $n$ for hyperbolic deformed defects, $\phi^{(n)}(s)$ (dashed red lines) and $\psi^{(n)}(s)$ (solid black lines), obtained from the $3$-cyclic deformation chain.
The topological charge for the primitive defect, $\chi$, is equal to $2\pi$.}
\label{Fig03carga}
\end{figure}

With the visual support from Fig. \ref{Fig03carga} one can see that there is a nearly constant topological charge for $n \neq 1$ and this value goes to zero in a strictly narrow region around $n = 1$. There is an anti-kink for $n\lesssim-1$ that continuously changes into a usual lump, at $n=1$, and then continuously changes into a kink for $n\gtrsim 3$.

Turning to the potential $V(\phi)$, it engenders a set of critical points, $\phi^{(n)0}_{\pm}$, that correspond to the asymptotic values of the kink-like solution, namely,
\begin{eqnarray}
\phi^{(n)0}_{\pm} &=& \phi^{(n)}(s\rightarrow \pm\infty) = \pm\left(\arctan\left[\sinh(n\pi)\right]+\arctan\left[\sech(\pi)\sinh[(1-n)\pi]\right]\right),
\end{eqnarray}
such that the topological charges in terms of $n$ are given by
\begin{equation}
Q^{\phi}_{(n)}=2\left\vert\arctan\left[\sinh(n\pi)\right]+\arctan\left[\sech(\pi)\sinh[(1-n)\pi]\right]\right\vert,
\label{cargaphi3hip}
\end{equation}
as it can be depicted in Fig.~\ref{Fig03carga}. One can notice that kink-like defects, $\phi^{(n)}$, appear for a certain interval $-2\lesssim n \lesssim 4$. Note that the charges given in Eq.~(\ref{cargaphi3hip}) are due to the topological defects for this interval, that includes the $n$ values used to calculate the kinks $\phi^{(n)}$ shown in Fig.~\ref{Fig03A}.
For $n\lesssim-2$ and $n\gtrsim4$, one can see through Fig.~\ref{Fig03carga} that $Q^{\phi}_{(n)}$ are highly suppressed, and therefore make $\phi^{(n)}$ behaving like lump solutions.

Performing a numerical integration, both hyperbolic mass dependencies on $n$, $M^{\psi}_{(n)}$ and $M^{\phi}_{(n)}$, can be obtained and are depicted in Fig.~\ref{Fig03C}.

\subsection{Trigonometric deformation}

Let us now turn to a set of trigonometric auxiliary functions described by
\begin{eqnarray}
\phi^{(n)}_{\chi} =  \cos[(\chi-(n-1)\pi)],\nonumber\\
\psi^{(n)}_{\chi} = -\sin[(\chi-(n-1)\pi)],
\label{hyp03B}
\end{eqnarray}
which upon straightforward integrations lead to
\begin{eqnarray}
\label{defdeformadostrig}
\phi(\chi,n) &=& \sin[\chi - (n - 1) \pi],\nonumber\\
\psi(\chi,n) &=& \cos[\chi - (n - 1) \pi] - \cos(\pi),
\label{hyp3B}
\end{eqnarray}
again, with constants and overall signs chosen to fit ordinary values for the asymptotic limits.
After simple mathematical manipulations involving the ordinary trigonometric fundamental relation $
\sin{(n\,\chi)}^2 + \cos{(n\,\chi)}^2 = 1
$ and relations from Eq.~(\ref{topo023B}), one recurrently identifies the closure relation
\begin{eqnarray}
w_{\chi}^{2}
&=& w_{\chi}^{2}\left[\sin{(n\,\chi)}^2 + \cos{(n\,\chi)}^2\right]\nonumber\\
&=& w_{\chi}^{2}\left[\psi^{(n)\,2}_{\chi} + \phi^{(n)\,2}_{\chi}\right]\nonumber\\
&=& z_{\psi}^{2} + y_{\phi}^{2}.
\label{hyp013B}
\end{eqnarray}

For $\chi(s)$ from Eq.~(\ref{solsine}), the BPS deformed functions are obtained as
\begin{eqnarray}
y_{\phi}=\omega_{\chi} \cos[\chi-(n-1)\pi], \nonumber\\
z_{\psi}=-\omega_{\chi} \sin[\chi-(n-1)\pi],
\label{hyp003B}
\end{eqnarray}
$\chi(s)$ into Eq.~(\ref{hyp3B}).

In this case, the potentials, $W(\psi)$ and $V(\phi)$ generate lump-like defects for all the solutions.
The second column of Fig.~\ref{Fig03A} shows the profile of defects and their corresponding energy densities, $\rho$'s, for $\chi(s)$, and for the corresponding deformed defects, $\phi^{(n)}(s)$, with $\rho(\phi^{(n)}(s))$, and $\psi^{(n)}(s)$, with $\rho(\psi^{(n)}(s))$.
We have set $n= 1 - 0.05k$ with $k\in[-2,2]$ in order to depict the analytical dependency on $n$. The defects correspond to lump-like structures given that the topological charges of $\psi$ and $\phi$ found to be equal to $0$.

The corresponding BPS potentials, $W(\psi)$ and $V(\phi)$, for the same set of $n$ parameters are depicted from the second column of Fig.~\ref{Fig03B}.

The lump-like solutions $\psi^{(n)}$ are obtained from a potential $W(\psi)$ that engenders a set of boundary points,  $\psi^{(n)0}_{\pm}$, corresponding to the asymptotic values of the lump-like solution, i. e.
\begin{equation}
\psi^{(n)0}_{\pm} = \psi^{(n)}(s\rightarrow \pm \infty) = 1 + \cos\left(\frac{n}{\pi}\right),
\end{equation}
where it is evinced that the topological charge is zero.
In this case, the corresponding total energy of the localized solution is given by
\begin{equation}
M^{\psi}_{(n)} = 4 + \frac{4}{15}\cos(2 n \pi).
\end{equation}

The potential $V(\phi)$ engenders a set of boundary points, $\phi^{(n)0}_{\pm}$, that correspond to the asymptotic values of the lump-like solution, i. e.
\begin{equation}
\phi^{(n)0}_{\pm} = \phi^{(n)}(s\rightarrow \pm \infty) = -\sin\left(\frac{n}{\pi}\right),
\end{equation}
from which one notices that the topological charge is zero.
Hereon one also notices another peculiar feature: a kind of, at least visually, {\em pseudo}kink behavior exhibited by $\phi^{(n)}$ (red lines), as depicted in the second column of Fig.~\ref{Fig03A}.
In spite of being a lump solution, it has two inflection points instead of the expected one.
The localized energy obtained analytically as function of the free parameter $n$ is expressed by
\begin{eqnarray}
M^{\phi}_{(n)} = 4 - \frac{4}{15}\cos(2 n \pi).
\end{eqnarray}

\begin{figure}[h!]
\centering
\includegraphics[scale=0.6]{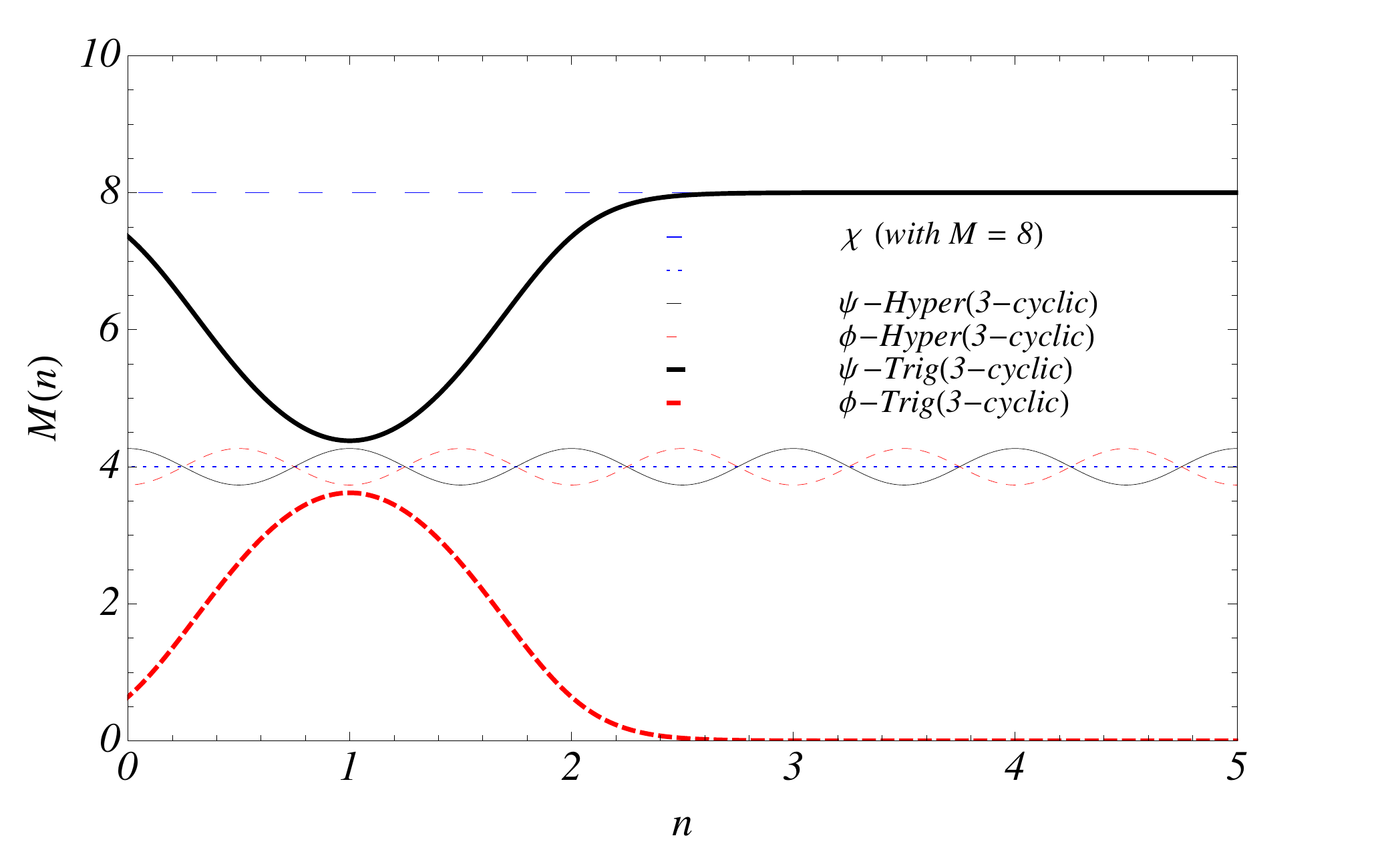}
\caption{(Color online) Total energy of localized solutions (topological masses), $M_{(n)}$, as function of the parameter $n$ for hyperbolic (thick lines) and trigonometric (thin lines) $4$-cyclic deformation chains.
Results are for $M^{\phi}_{(n)}$ (dashed red lines) and $M^{\psi}_{(n)}$ (solid black lines).
The topological mass for the primitive defect, $\chi$, is given by $M^{\chi} = 8$.}
\label{Fig03C}
\end{figure}

Fig.~\ref{Fig03C} summarizes the results for the localized energies analytically expressed in terms of $n$, from which the constraint equation obtained from Eqs.~(\ref{hyp013}) and (\ref{hyp013B}) results into
\begin{equation}
M^{\psi}_{(n)} + M^{\phi}_{(n)}  = M^{\chi} = 8,
\end{equation}
which is analytically verified for both hyperbolic and trigonometric $3$-cyclic deformation chains.
Fig \ref{Fig03C} also attest that for both hyperbolic and trigonometric deformations the mass summation of $3$-cyclically deformed defects results into the mass of the primitive SG defect, $M^{\chi}$.

\section{$4$-cyclic SG models}

The analysis performed in the previous section can be straightforwardly extended to a $4$-cyclic deformation chain scenario.
Also in this case, a modified constraint relation involving the topological masses is obtained, and analogous results involving charge varying topological scenarios can be reissued.

First of all, an additional scalar field, $\varphi$, has to be considered in order to complete the $4$-cyclic chain supported by the primitive triggering SG defect, $\chi$.
The coupled system described by Eqs.~(\ref{topo023}-\ref{topo023B}) can thus be extended to
\begin{eqnarray}
w_{\chi}    &=& \frac{d \chi}{d s} = x_{\varphi} \chi_{\varphi} = y_{\phi} \chi_{\phi} = z_{\psi} \chi_{\psi},
\label{topo024}
\end{eqnarray}
where
\begin{eqnarray}
x_{\varphi} &=& w_{\chi} \varphi_{\chi}, \nonumber\\
y_{\phi}    &=& w_{\chi} \phi_{\chi}, \nonumber\\
z_{\psi}    &=& w_{\chi} \psi_{\chi},
\label{topo024B}
\end{eqnarray}
and the cyclically derived BPS potentials belonging to the $4$-cyclic deformation chain will be given by
\begin{equation}
T(\chi) = \frac{1}{2} w_{\chi}^2 ~\rightleftarrows ~
U(\varphi) = \frac{1}{2} x_{\varphi}^2 ~\rightleftarrows ~
V(\phi) = \frac{1}{2} y_{\phi}^2 ~\rightleftarrows ~
W(\psi) = \frac{1}{2} z_{\psi}^2 ~\rightleftarrows ~
T(\chi) = \frac{1}{2} w_{\chi}^2.
\label{topo024C}
\end{equation}

The four deformation functions shall naturally follow the chain rule given by
\begin{equation}
\psi_{\phi} \phi_{\varphi} \varphi_{\chi} \chi_{\psi}= 1.
\label{topo034}
\end{equation}

\subsection{Hyperbolic Deformation}

As one has one more cycle, the set of auxiliary functions is now described by
\begin{eqnarray}
\psi^{(n)}_{\chi} &=& -\tanh[\chi-(n-1)\pi], \nonumber\\
\phi^{(n)}_{\chi} &=& -\tanh{[\chi-(n-1)\pi]}\sech{[\chi-(n-1)\pi]}, \nonumber\\
\varphi^{(n)}_{\chi} &=& \sech{[\chi-(n-1)\pi]}^{2},
\label{hyp00}
\end{eqnarray}
where the function $w_{\chi}$ substituted into Eq.(\ref{topo024B}) completes the $4$-cyclic chain.
Performing the corresponding integrations, one gets the defects
\begin{eqnarray}
\psi^{(n)}(\chi)     &=& -\ln{\left[\frac{\cosh{[\chi-(n-1)\pi]}}{\cosh{(\pi)}}\right]}, \nonumber\\
\phi^{(n)}(\chi)     &=& \sech{[\chi-(n-1)\pi]} -\sech{(\pi)}, \nonumber\\
\varphi^{(n)}(\chi) &=& \tanh{[\chi-(n-1)\pi]},
\label{hyp}
\end{eqnarray}
which fit the ordinary values for the asymptotic limits.
Following the same simple mathematical manipulations, one easily identifies the equality,
\begin{eqnarray}
w_{\chi}^{2}
&=& w_{\chi}^{2}\left[\tanh{(n\,\chi)}^2 + \sech{(n\,\chi)}^2\left(\tanh{(n\,\chi)}^2 + \sech{(n\,\chi)}^2\right)\right]\nonumber\\
&=& w_{\chi}^{2}\left[\psi^{(n)\,2}_{\chi} + \phi^{(n)\,2}_{\chi} + \varphi^{(n)\,2}_{\chi}\right]\nonumber\\
&=& z_{\psi}^{2} + y_{\phi}^{2} + x_{\varphi}^{2},
\label{hyp1}
\end{eqnarray}
which constrains the values for the topological masses of the hyperbolically deformed defects.
The other corresponding BPS deformed functions are obtained as
\begin{eqnarray}
x_{\varphi}(s) &=& \omega_{\chi}\sech{[\chi-(n-1)\pi]}^{2} ,\nonumber\\
y_{\phi}(s)    &=& -\omega_{\chi}\tanh{[\chi-(n-1)\pi]}\sech{[\chi-(n-1)\pi]},\nonumber\\
z_{\psi}(s)    &=& -\omega_{\chi} \tanh[\chi-(n-1)\pi].
\label{kinksmmm}
\end{eqnarray}

\begin{figure}[h!]
\begin{center}
\hspace{-1 cm}
\includegraphics[scale=0.5]{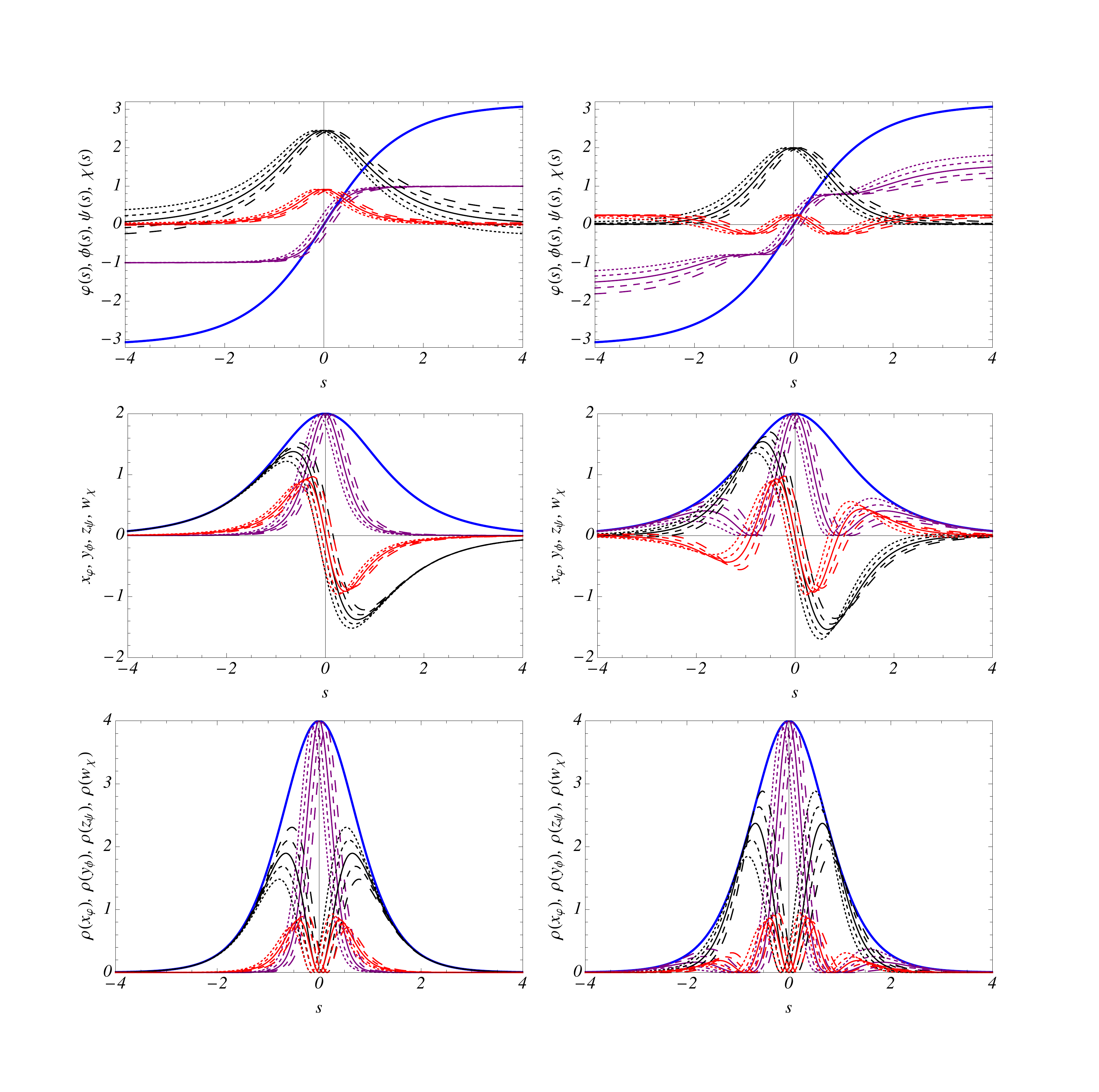}
\end{center}\caption{(Color online) The $4$-cyclically deformed defects systematically obtained from hyperbolic (first column) and trigonometric (second column) deformation functions.
Results are for the primitive SG kink solution, $\chi(s)$ (thick blue line), and for deformed defects, $\phi^{(n)}(s)$ (red lines), $\psi^{(n)}(s)$ (black lines) and $\varphi^{(n)}(s)$ (purple lines).
The hyperbolic deformation chain leads to effective kink-like defects with non-vanishing topological charges (with the exception of $n = 1$).
The trigonometric deformation chain leads to effective lump-like structures with vanishing topological charges (with the exception of $\varphi^{(n)}$ solutions).
On the first row one has the defect profile, on the second row one has the BPS function and on the last row one has the energy density profile.
The values of the free parameter $n$ are picked-up from the interval between $0.9$ (dotted lines) and $1.1$ (long dashed lines) as in Fig.~(\ref{Fig03A}).
Solid lines correspond to $n = 1$.}
\label{Fig04A}
\end{figure}

The first column of Fig.~\ref{Fig04A} shows the results for the primitive defects: $\chi(s)$, $w_{\chi}(s) = d\chi/ds$, and $\rho(\chi(s))$; and for the corresponding deformed defects: $\phi^{(n)}(s)$, $y_{\phi}(s) = d\phi^{(n)}/ds$, and $\rho(\phi^{(n)}(s))$; $\varphi^{(n)}(s)$, $x_{\varphi}(s) = d\varphi^{(n)}/ds$, and $\rho(\varphi^{(n)}(s))$; and $\psi^{(n)}(s)$, $z_{\psi}(s) = d\psi^{(n)}/ds$, and $\rho(\psi^{(n)}(s))$ (from hyperbolic deformation functions).
We have set $n = 1 - 0.05k$ with $k\in[-2,2]$, in order to depict the analytical dependence on the free parameter $n$.

The potential $W(\psi)$ produces a set of critical points, $\psi^{(n)0}_{\pm}$,  corresponding to the asymptotic values of the kink-like solution, namely
\begin{eqnarray}
\psi^{(n)0}_{+} &=& \psi^{(n)}(s\rightarrow +\infty) = -\ln[\sech(\pi)\cosh[(n-2)\pi]],\nonumber\\
\psi^{(n)0}_{-} &=& \psi^{(n)}(s\rightarrow -\infty) = -\ln[\sech(\pi)\cosh[n\pi]].
\end{eqnarray}

The critical points $\phi^{(n)0}_{\pm}$ engendered by the potential $V(\phi)$, which correspond to the asymptotic values of the kink-like solution, are given by:
\begin{eqnarray}
\phi^{(n)0}_{+} &=& \phi^{(n)}(s\rightarrow +\infty) = \sech[(n-2)\pi] -\sech[\pi],\nonumber\\
\phi^{(n)0}_{-} &=& \phi^{(n)}(s\rightarrow -\infty) = \sech[n\pi]-\sech[\pi].
\end{eqnarray}

Finally, the potential $U(\varphi)$ engenders a set of critical points, $\varphi^{(n)0}_{\pm}$, that correspond to the asymptotic values of the kink-like solution, i. e.
\begin{eqnarray}
\varphi^{(n)0}_{+} &=& \varphi^{(n)}(s\rightarrow +\infty) = -\tanh[(n-2)\pi],\nonumber\\
\varphi^{(n)0}_{-} &=& \varphi^{(n)}(s\rightarrow -\infty) = -\tanh[n\pi].
\end{eqnarray}

Hence all the topological charges for the deformed defects obtained from the hyperbolically deformation can thus be written as
\begin{eqnarray}
Q^{\psi}_{(n)}&=&\left\vert\ln\left[\frac{\sech[(n-2)\pi]}{\sech[n\pi]}\right]\right\vert,\nonumber\\
Q^{\phi}_{(n)}&=&\left\vert\sech[(n-2)\pi] - \sech[n\pi]\right\vert\nonumber\\
Q^{\varphi}_{(n)}&=&\left\vert\tanh[(n-2)\pi] - \tanh[n\pi]\right\vert\,.
\end{eqnarray}
Their behavior can be depicted in the Fig.~\ref{Figcarga4}.

Once again the deformed defects $\psi^{(n)}$ show {\em pseudo}lump features (black lines in the first column of Fig.~\ref{Fig04A}), that are kink solutions indeed.
Otherwise, one has $\phi^{(n)}\sim$ constant for $n\lesssim2$ and $n\gtrsim 4$. In the interval between these values there is the formation of an anti-kink for $n \sim 0$, that continuously transforms into a lump, at $n = 1$, and keep changing to form a kink at $n \sim 2$. Such a behavior is in complete agreement with that depicted in Fig.~\ref{Figcarga4} (c. f. red line results).

The topological charges for the kink deformed defects obtained from hyperbolic deformations are computed through numerical integration.
Their behavior can be depicted in Fig.~\ref{Figcarga4}.
Again, the charge varying topological behavior shows that the dependence on the phenomenological parameter $n$ creates a novel scenario where a smooth transition from topological to asymptotically non-topological scenarios is possible.

In this case, the topological masses are obtained from numerical integrals and they can be seen in Fig. \ref{Fig04C}.

\subsection{Trigonometric Deformation}

Turning again to the set of auxiliary trigonometric functions described by
\begin{eqnarray}
\psi^{(n)}_{\chi} &=& -\sin{[\chi-(n-1)\pi]}, \nonumber\\
\phi^{(n)}_{\chi} &=&  -\cos{[\chi-(n-1)\pi]}\sin{[\chi-(n-1)\pi]},\nonumber\\
\varphi^{(n)}_{\chi} &=&  \cos{[\chi-(n-1)\pi]}^2,
\label{trig0}
\end{eqnarray}
upon straightforward integrations, one obtains
\begin{eqnarray}
\psi^{(n)}(\chi) &=& \cos{[\chi-(n-1)\pi]} - \cos{(\pi)}, \nonumber\\
\phi^{(n)}(\chi) &=& \frac{1}{4}\left[\cos{[2(\chi-(n-1)\pi))]}\right], \nonumber\\
\varphi^{(n)}(\chi) &=&  \frac{[\chi-(n-1)\pi]}{2}+\frac{1}{4}\sin{[2(\chi-(n-1)\pi)]},
\label{trig}
\end{eqnarray}
with constants chosen to follow the same asymptotic limit criteria.
By following the same simple mathematical manipulations involving the fundamental relation between $\sin{(n\,\chi)}$ and $\cos{(n\,\chi)}$, one easily identifies that
\begin{eqnarray}
w_{\chi}^{2}
&=& w_{\chi}^{2}\left[\sin{(n\,\chi)}^2 + \cos{(n\,\chi)}^2\left(\sin{(n\,\chi)}^2 + \cos{(n\,\chi)}^2\right)\right]\nonumber\\
&=& w_{\chi}^{2}\left[\psi^{(n)\,2}_{\chi} + \phi^{(n)\,2}_{\chi} + \varphi^{(n)\,2}_{\chi}\right]\nonumber\\
&=& z_{\psi}^{2} + y_{\phi}^{2} + x_{\varphi}^{2}
\label{trig1}
\end{eqnarray}
constrains the values for the topological masses of the trigonometrically deformed defects.

Eqs.~(\ref{trig0}) lead to the other BPS deformed lump-like solutions,
\begin{eqnarray}
x_{\varphi}(s) &=& \omega_{\chi}\cos{[\chi-(n-1)\pi]}^2,\nonumber\\
y_{\phi}(s)    &=& -\omega_{\chi}\sin{[\chi-(n-1)\pi]}\cos{[\chi-(n-1)\pi]},\nonumber\\
z_{\psi}(s)    &=& -\omega_{\chi}\sin{[\chi-(n-1)\pi]},
\label{kinkstrig}
\end{eqnarray}
once again one has assumed $\chi$ from Eq.(\ref{solsine}).

The potential $U(\varphi)$ generates kinks and the potentials $V(\phi)$ and $W(\psi)$ generate lump-like defects. The second column of Fig.~\ref{Fig04A} shows the results for the primitive defects: $\chi(s)$, $w_{\chi}(s) = d\chi/ds$, and $\rho(\chi(s))$; and for the corresponding deformed defects: $\phi^{(n)}(s)$, $y_{\phi}(s) = d\phi^{(n)}/ds$, and $\rho(\phi^{(n)}(s))$; $\varphi^{(n)}(s)$, $x_{\varphi}(s) = d\varphi^{(n)}/ds$, and $\rho(\varphi^{(n)}(s))$; and $\psi^{(n)}(s)$, $z_{\psi}(s) = d\psi^{(n)}/ds$, and $\rho(\psi^{(n)}(s))$ (from trigonometric deformation functions).
Again we have set $n = 1 - 0.05k$ with $k\in[-2,2]$.

\begin{figure}[h!]
\centering
\includegraphics[scale=0.7]{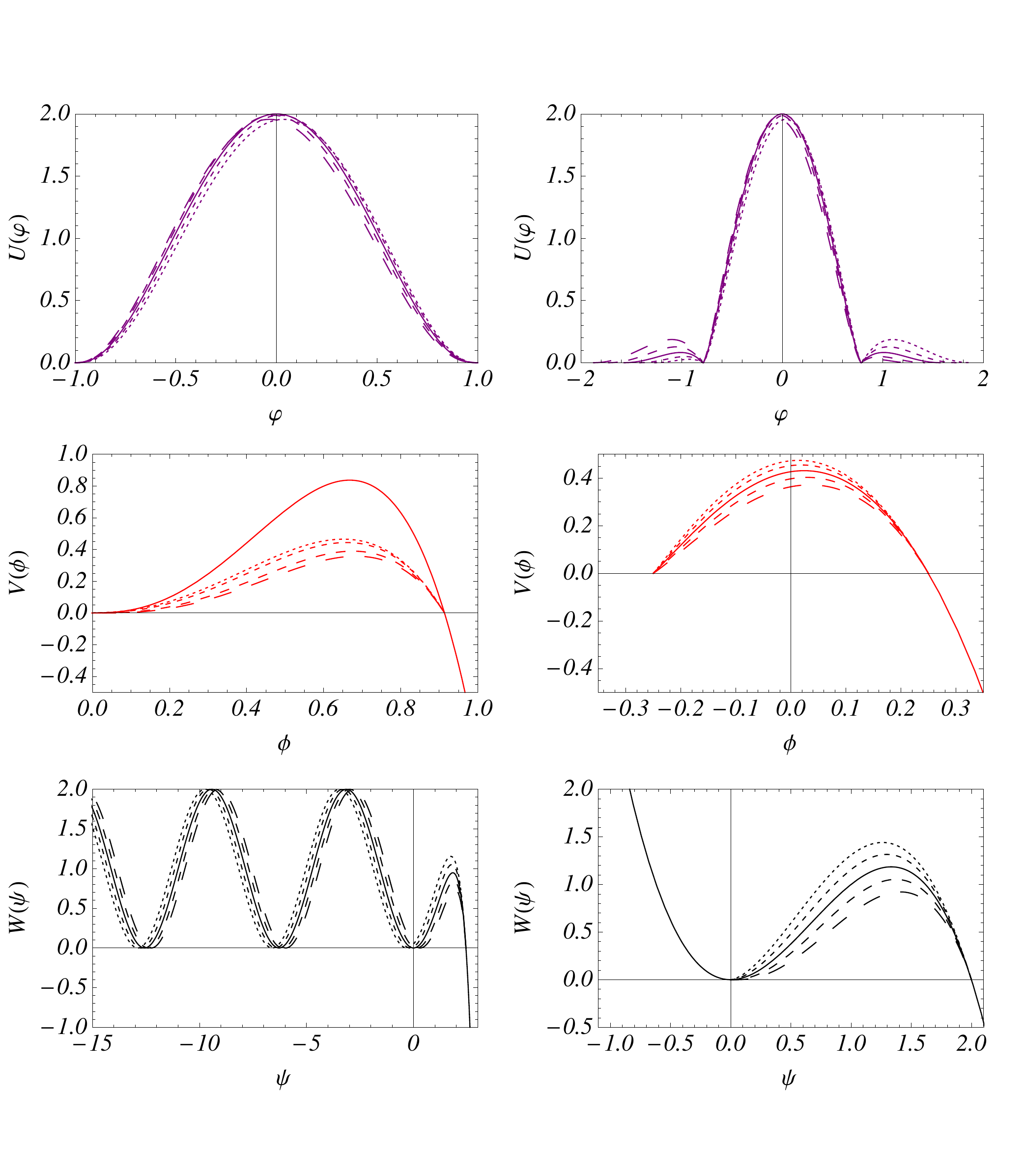}
\caption{(Color online) Deformed potentials, $V(\phi)$, $W(\psi)$ and $U(\varphi)$, obtained from the SG model, for hyperbolic (first column) and trigonometric (second column) $4$-cyclic deformation chains.
We have selected the same values of $n$ and we have set the {\em color scheme} correspondence  from Fig.~\ref{Fig04A}.}
\label{Fig04B}
\end{figure}

\begin{figure}[h!]
\centering
\includegraphics[scale=0.7]{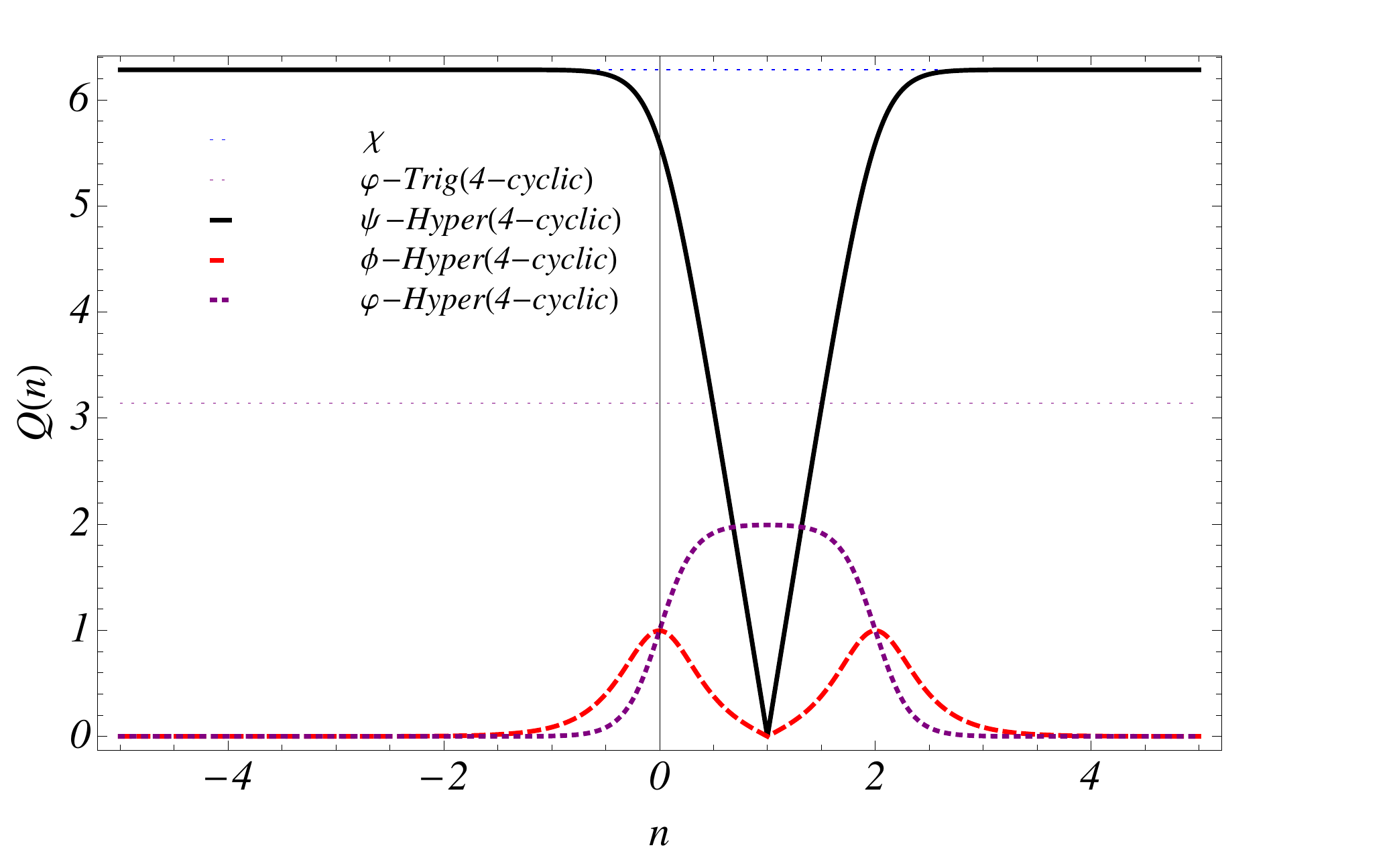}
\caption{(Color online) Topological charges, $Q_{(n)}$, for kink solutions as function of the parameter $n$ for hyperbolic deformed defects, $\phi^{(n)}(s)$ (dashed red lines), $\psi^{(n)}(s)$ (solid black lines), and $\varphi^{(n)}(s)$ (dotted purple lines), obtained from the $4$-cyclic deformation chain.
The topological charge for the primitive defect, $\chi$, is equal to $2\pi$.}
\label{Figcarga4}
\end{figure}

\begin{figure}[h!]
\centering
\includegraphics[scale=0.7]{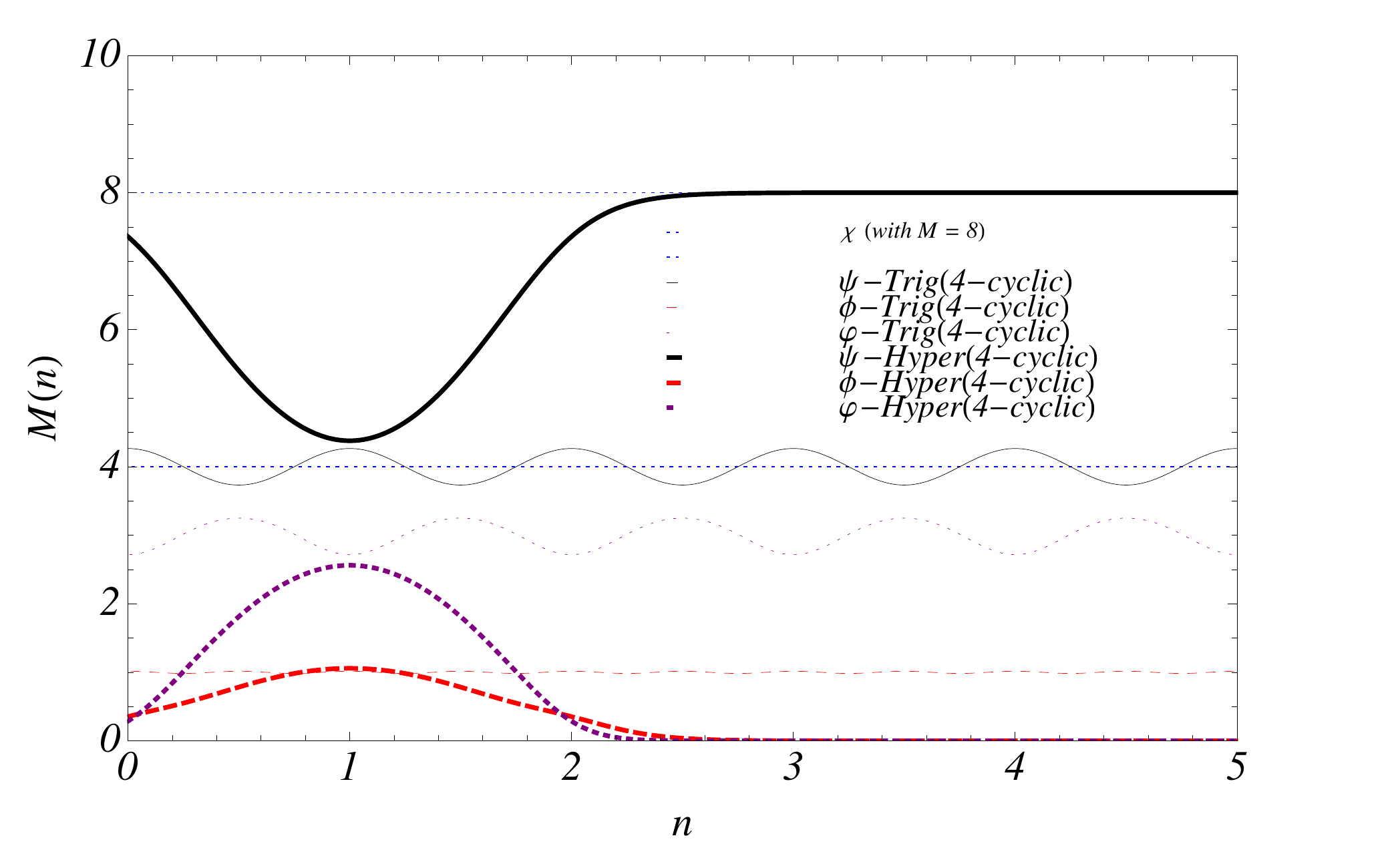}
\caption{(Color online) Total energy of localized solutions (topological masses, $M_{(n)}$) as function of the parameter $n$ for hyperbolic (thick lines) and trigonometric (thin lines) $4$-cyclic deformation chains.
Results are $M^{\phi}_{(n)}$ (dashed red lines), $M^{\psi}_{(n)}$ (solid black lines), and  $M^{\varphi}_{(n)}$ (dotted purple lines).
The topological mass for the primitive defect, $\chi$, is given by $M^{\chi} = 8$.}
\label{Fig04C}
\end{figure}

The corresponding BPS potentials, $W(\psi)$, $V(\phi)$ and $U(\varphi)$, for the same set of $n$ values, are depicted from the second column of Fig.~\ref{Fig04B}.

The potentials $W(\psi)$ and $V(\phi)$ arouse a set of boundary points that correlate to the asymptotic values given by
\begin{equation}
\psi^{(n)0}_{\pm} = \psi^{(n)}(s\rightarrow \pm \infty) = 1 + \cos\left(n\pi\right),
\end{equation}
and
\begin{equation}
\phi^{(n)0}_{\pm} = \phi^{(n)}(s\rightarrow \pm \infty) = \frac{1}{4}\cos\left(2 n\pi\right)\,.
\end{equation}
Consequently, one notices that the corresponding topological charges are null and are correlated to lump-like solutions, even though one sees that for $\phi^{(n)}$ they are unusual $3$-inflection point lumps.

Otherwise, the potential $U(\varphi)$ engenders a set of critical points, $\varphi^{(n)0}_{\pm}$, which correspond to the asymptotic values of a kink-like solution,
\begin{eqnarray}
\varphi^{(n)0}_{+} &=& \varphi^{(n)}(s\rightarrow +\infty) = \pi-\frac{n\pi}{2}-\frac{1}{4}\sin(2n\pi),\nonumber\\
\varphi^{(n)0}_{-} &=& \varphi^{(n)}(s\rightarrow -\infty) = 0 -\frac{n\pi}{2}-\frac{1}{4}\sin(2n\pi),
\end{eqnarray}
such that the topological charge for $\varphi^{(n)}$ is simply given by
\begin{equation}
Q_{(n)}^{\varphi} = \pi.
\label{kinksstrig}
\end{equation}
This charge is depicted in Fig.~\ref{Figcarga4}.
Finally, the topological masses follow the same previously obtained chain property such that
\begin{eqnarray}
M^{\psi}_{(n)} &=& 4 + \frac{4}{15}\cos{(2n\pi)},\nonumber\\
M^{\phi}_{(n)} &=& 1 + \frac{1}{63}\cos{(4n\pi)},\nonumber\\
M^{\varphi}_{(n)} &=& 3-\frac{4}{15}\cos{(2n\pi)}-\frac{1}{63}\cos{(4n\pi)},
\label{masstrig}
\end{eqnarray}
as shown in Fig.~\ref{Fig04C}, from which one obtains
\begin{equation}
M^{\psi}_{(n)} + M^{\phi}_{(n)} +M^{\varphi}_{(n)} = M^{\chi} = 8.
\end{equation}

One can see that the kink given by $\varphi^{(n)}$ seems to support some exotic two-kink solutions \cite{bazeia0sinegordon}.
This kink-like structure is seen not only in \cite{bazeia0sinegordon}, but also in a braneworld scenario \cite{Bas05} where one sees the appearance of a new phase between the two interfaces that supports some brane internal structure.

\section{Conclusions}

In this manuscript, a the systematic procedure supported by cyclic deformation chains has been reissued in order to investigate novel models derived from the SG theory.
It has been applied a set of rules for constructing constraint relations involving topological masses and for obtaining analytical expressions for the topological charge of the deformed defects cyclically generated.
In particular, it has been ratified the possibility of re-obtaining primitive SG defect structures through a regenerative and unidirectional sequence of deformation operations, via hyperbolic and trigonometric functions, which is typical from an $N\hspace{-.1 cm}+\hspace{-.1 cm}2$-cyclic deformation chain.

An additional issue has concerned the obtaining of our deformed models.
In general, the presence of nonlinearities in the differential equations has made it difficult to find analytic solutions and to compute the quantifiers as the topological mass and the topological charge.
Therefore, the solutions here obtained are relevant {\em per se} since they correspond to a class of no elementary solutions of such equations.

The main effect produced by cyclic deformations that has been evinced through our results is concerned with the possibility of obtaining typical deformed SG solutions as nonmonotonic kinks or even as $N\hspace{-.1 cm}+\hspace{-.1 cm}2$-inflection point lumps.
The deformed SG models here obtained indeed exhibit a remarkable and distinctive feature: the possibility of smooth and continuos transitions between topological ($Q \neq 0$) and (at least asymptotically) non-topological ($Q = 0$) scenarios, and analogous reversal transitions,  depending on a phenomenological degree of freedom parameterized by $n$.
Furthermore, for a particular structure belonging to the $4$-cyclic chain, one has identified a deformed defect (c. f. the 
$\varphi(s)$ defect) roughly behaving like a double kink solution, which contains a great wealth of detail.

Our analytical solutions can also be used for simplifying the calculation of the internal modes of the theory. 
They may support the formation of two-kink soliton pairs in perturbed sine-Gordon models due to kink internal-mode instabilities \cite{PRE2012}, or even the existence of internal modes of deformed kinks as identified in \cite{PRE2000,PRE2000B}.

Through systematic defect deformation mechanisms \cite{Bas01,AlexRoldao} one can, for instance, identify whether generic solutions, $\varphi(s)$, are stable under time-dependent small perturbations \cite{Col85}.
For scalar field potentials, $U(\varphi)$, that engender kink- and lump-like structures \cite{Bas01,Bas02}, the BPS first-order framework \cite{BPS,BPS2,BookB,Bas01} sets 
\begin{equation}
U(\varphi) = z_{\varphi}^{2}/2,~\mbox{with}~z_{\varphi} = dz/d\varphi = d\varphi/ds = \varphi^{\prime},
\label{BPS}
\end{equation}
and, obtained from perturbative corrections of the field equations of motion, a Schr\"odinger like equation is written as
\begin{equation}
\left(- d^2/ds^2 + \varphi^{\prime\prime\prime}/\varphi^{\prime}\right)\psi_{n} = \omega_n^2\,\psi_{n}.
\label{SC2}
\end{equation}
Simple manipulations lead to a zero-mode, $\psi_{0}$ (i. e. when $\omega_0 = 0$), given by $\psi_{0} \propto \varphi^{\prime}$, which corresponds to the quantum ground state when $\varphi(s)$ is a kink for which $\varphi^{\prime}$ has no zeros (nodes).
In this case, there would be no instability from modes with $\omega_n^2 < 0$.
To summarize, it has been shown that the zero-mode stable vacuum (ground state) solutions supported by a topological defect can be converted into an unstable (tachyonic) quantum state supported by the corresponding deformed topological defect.
As one knows, kinks are topological monotonic solutions for which the derivatives with respect to the position coordinate, $s$, have no nodes (zeros).
In that case they support stable zero-mode (ground state) solutions of the Schr\"odinger-like equation for perturbations (c. f. Eq.~(\ref{SC2})). 
Analogously, lumps are nontopological nonmonotonic solutions for which the derivatives with respect to $s$ have (at least) one node (zero). Therefore, lumps support unstable $n$-level excited state solutions of the Schr\"odinger-like equation, where $n$ is the number of zeros of $\varphi^{\prime} = d\varphi/ds$.
The same criterium is adjudicated to nonmonotonic kinks (and lumps), which hence support unstable solutions.

Finally, in a braneworld scenario, such a double kink structure could be associated with models described by potentials that drive the system to support thick brane solutions that engender internal structures unveiling the presence of a novel internal phase in the brane.
Our solutions are therefore important not only in applied physics scenarios, namely in solid state physics and optics, but also in cosmological scenarios supported by braneworld dynamics: the analytical structures here described can also be investigated in some typical scenarios of structure formation \cite{0001C,0002C,0003C,0003B}, Q-balls \cite{0004C,0004B}, spinor effective mass generation induced by some symmetry breaking mechanism \cite{Ber12}, tachyonic branes \cite{0005,0006,0007,Bertolami}, and in braneworld scenarios with a single extra-dimension of infinity extent \cite{Bas05,Bas05B,0008,0009}.

{\em Acknowledgments - This work was supported by the Brazilian Agency CNPq (grant 440446/2014-7, grant 300809/2013-1, grant 303027/2012-6 and grant 473326/2013-2).}

\renewcommand{\baselinestretch}{1.0}

\end{document}